# Replication of proto-RNAs sustained by ligase-helicase cycle in oligomer world

Daisuke Sato, Osamu Narikiyo*

*Department of Physics, Kyushu University, Fukuoka 812-8581, Japan*

ABSTRACT

A mechanism of the replication of proto-RNAs in oligomer world is proposed. The replication is carried out by a minimum cycle which is sustained by a ligase and a helicase. We expect that such a cycle actually worked in the primordial soup and can be constructed in vitro. By computer simulation the products of the replication acquires diversity and complexity. Such diversity and complexity are the bases of the evolution.

\* Corresponding author.
   *E-mail address:* narikiyo@phys.kyushu-u.ac.jp (O. Narikiyo)

# 1. Introduction

On the origin of life there are three major hypotheses: RNA-world hypothesis (Atkins et al., 2011), protein-world hypothesis (Shapiro, 2007) and lipid-world hypothesis (Luisi, 2006). However, to meet all of the conditions, replication, metabolism and compartment (Szathmary, 2006), for the establishment of a proto-cell it is natural to employ all of the proto-elements, proto-RNA, proto-protein and proto-membrane, for the establishment. Although the lipid-membrane is not modeled explicitly, all of the proto-elements are taken into account in the oligomer-world hypothesis. This hypothesis is advocated by Shimizu (Shimizu, 1996) and we have constructed a proto-cell in oligomer world by a computer simulation (Nishio and Narikiyo, 2013).

In our previous study based on the oligomer-world hypothesis (Nishio and Narikiyo, 2013) the process of the replication of proto-RNAs is not constructed explicitly but handled as the simulation rule. Here in this paper we will propose the way of the replication by a computer simulation. The elements of the simulation have been obtained already by experiments so that such a simulation is expected to be a simplified demonstration of what actually occurred in the primordial soup.

To construct the replication system in the oligomer world we need at least two types of oligomers, ligase and helicase. Here the ligase acts as the template of the replication. The ligase which we employ has been established in vitro (Lincoln and Joyce, 2009) as an oligo-nucleotide. It bridges two oligo-nucleotides and is much smaller in numbers of nucleotides than that of polymerase which generates poly-nucleotides from monomers of nucleotides. The bridged oligo-nucleotides are complementary to the ligase in the sense of the Watson-Crick pairing. The ligase is the template of the replication and the first digital machine in our oligomer world.

Unfortunately the ligase cannot work for the other oligo-nucleotides when the ligase and the bridged oligo-nucleotides are bound together by the Watson-Crick pairing interaction. In vitro they are separated by the heat cycle of the PCR and there is an argument that such a cycle might work near the spout of the hot water in the sea (Ricardo and Szostak, 2009). However, the time-scale of the water convection near the spout is unrelated to that of the ligation. It is desirable that the separation is carried out only when the ligation has finished. Moreover, the present-day cells employ a specific molecular motor, helicase, for the separation. Although the spout is suitable for the production of oligomers, it is natural to employ a primitive helicase because of the linkage with the present-day cells free from the spout. Fortunately some molecular motor consisting of mini proteins has been reported (Cordin et al., 2006; Patel and Donmez, 2006) to have a helicase function, while the present-day helicases are complex molecular motors.

Consequently the ligase and the helicase form a minimum cycle to replicate the template ligase. We expect that such a cycle actually worked in the primordial soup and can be constructed in vitro.

## 2. Model

The central element of our model is a ligase which is an oligo-nucleotide consisting of 20 nucleotide monomers. This ligase (L) models the real one (Lincoln and Joyce, 2009) in vitro. To have a ligation function it should have specific bases around the center, G-A-A-G from 9-th to 12-th base counted from its 5'-end. The functional part of the ligase, G-A-A-G, combines 5'-end and 3'-end of the substrates. The substrates of the ligation consist of 10 nucleotide monomers. Two of the substrate oligo-nucleotides are bridged by the ligase if they meet the following two conditions. (i) One substrate (S1) has the bases G-A at 5'-end and the other (S2) has G-A at 3'-end. Namely they also have the same bases, G-A-A-G, around the bridge as L's center when 5'-end and 3'-end are combined. To form a structure suitable for the ligation the ligase and the substrates are not complementary at the functional part as observed by the experiment (Lincoln and Joyce, 2009). (ii) The ligation proceeds only when the support of the other regions of the ligase and the substrates is strong enough. The strength of the support is quantified in our model by the number of complementary base-pairs between from 1-st to 6-th counted from L's 5'-end and from 1-st to 6-th counted from S1's 3'-end and between from 1-st to 6-th counted from L's 3'-end and from 1-st to 6-th counted from S2's 5'-end. If more than 4 Watson-Crick pairings are possible both for between L and S1 and between L and S2 at these region, then the support is regarded as strong enough. Both two conditions mimics the real ligation (Lincoln and Joyce, 2009) in vitro. There is no limitation for the bases at 7-th or 8-th from 5'-end or 3'-end. These conditions are illustrated in Fig. 1.

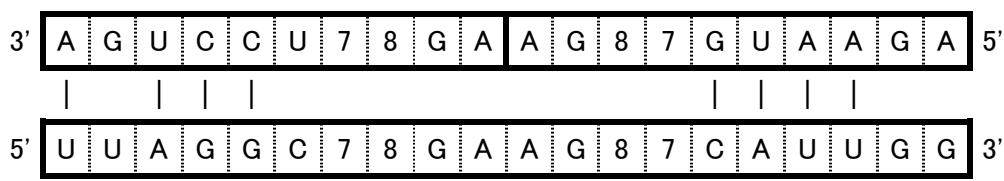

**Fig. 1.** An example for which the conditions (i) and (ii) are satisfied. The upper row represents S1 and S2. The lower row represents L. The vertical lines between rows represent the Watson-Crick pairings.

The ligase L loses the activity to the other oligo-nucleotides when S1 and S2 are bound. To separate L and S1-S2 we employ the helicase (H). H is a molecular motor and is assumed to consist of two domains, D1 and D2, of mini proteins in our model. Such a helicase is reported by experiments (Cordin et al., 2006; Patel and Donmez, 2006). In our simulation the detailed motion of H is not modeled but its activity is controlled by the numbers of effective domains, D1 and D2, where a pair of D1 and D2 forms a helicase. Our simulation cell contains random oligomers and the numbers of D1 and D2 are determined by the control parameter $D$ explained in the following. We have neglected the time for the binding between D1 and D2 and the time for the separation of L and S1-S2 for simplicity. Namely the formation of H and the motion of H, driven by proto-ATPs, are much quicker than the other activities, not driven by proto-ATPs, described in our model. Although we have neglected the time for the binding, we have retained a probability $p$ to success the binding as follows.

Our simulation starts from the situation where one ligase L is contained in the cell. The cell membrane is assumed to grow by a simple chemical reaction contained in the cell. Such a reaction is assumed in the models of chemoton (Ganti, 2003) or autopoiesis (Luisi, 2006). The membrane is assumed to be semi-permeable so that it keeps the ligase inside but takes in oligo-nucleotides and oligo-peptides, which are smaller than the ligase, from outside. The other materials, for example proto-ATPs necessary to drive helicases, are assumed to be supplied in the environment of the cell and to go through the membrane. We only manage oligo-nucleotides and oligo-peptides but do not manage these materials.

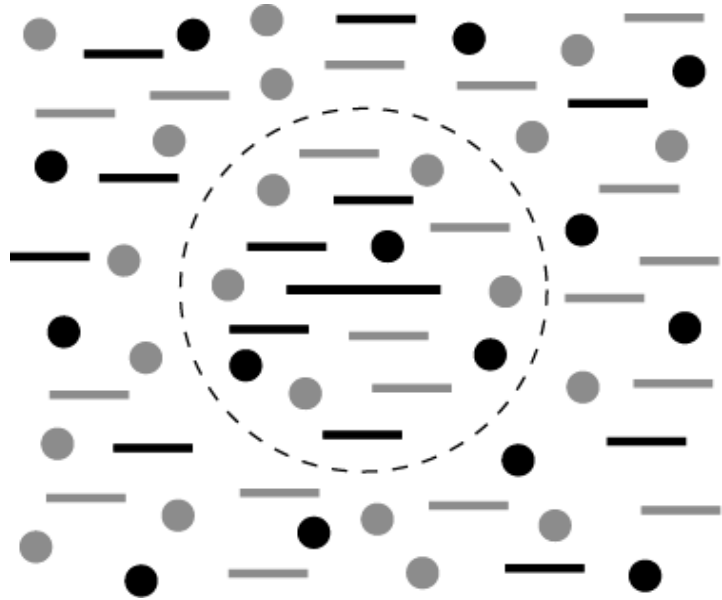

**Fig. 2.** Schematic representation of the initial cell in an environment. The broken line represents the semi-permeable membrane. Outer oligo-nucleotides (sticks) and oligo-peptides (disks) are supplied randomly and taken into the cell (the inside of the membrane) randomly according to the relation (1). One longer stick in the cell represents the ligase L. Oligo-nucleotides are classified into two categories; one (black stick) becomes a substrate, S1 or S2, of the ligation and the other

(gray stick) doesn't. Oligo-peptides are classified into two categories; one (black disk) becomes a domain of the helicase, D1 or D2, and the other (gray disk) doesn't.

The oligo-nucleotides and oligo-peptides are supplied in the environment and taken into the cell randomly. Every oligomer is assumed to be a random 10-mer. The accommodation number of oligomers, $n_N$ for oligo-nucleotides and $n_P$ for oligo-peptides, is determined as the largest number satisfying the relation

$$10n_N + 10n_P + 20n_L + 20n_H \leq V \tag{1}$$

where $n_L$ is the number of the ligase and $n_H$ the helicase. Here we have assumed that the volume of all the monomers, nucleotides and amino-acids, is unity for simplicity. The volume of the cell $V$ is estimated as that of the sphere $V = 4\pi r^3 / 3$ with the radius $r$. The radius is determined by the number of membrane molecules $n_M$ to satisfy $n_M = 4\pi r^2$. $n_L$ increases by one when the ligation occurs and the ligase and the substrates are separated. After the separation the product 20-mer of the ligation acts as a new ligase. $n_H$ increases by one when H is formed in the presence of the component domains D1 and D2. $n_N$ decreases by two when a new ligase is formed. $n_P$ decreases by two when a new helicase is formed. $n_N$ and $n_P$ increase according to (1) when the membrane expands. The ratio $R \equiv$ (the increase of $n_P$) / (the increase of $n_N + n_P$) at the expansion is given as a simulation parameter. At the same time helicase domains are introduced according to the control parameter $D \equiv$ (the increase in the number of the sum of D1 and D2) / (the increase of $n_P$) at the expansion. For simplicity we have neglected the difference in the number of D1 and the number of D2. If a pair of D1 and D2 exists in the cell, it is assumed to form a helicase by the probability $p$ which is a simulation parameter.

We have neglected the life-times of ligases and helicases. Namely we have assumed that they exist stably, once formed. All the oligomers supplied in the environment have been also assumed to be stable.

The number of membrane molecules $n_M$ increases from the initial number $n_{M0}$ up to $2n_{M0}$. The number increases by $\Delta n_M$ where we have assumed the existence of the chemical reaction mentioned above, in the cell, to produce the membrane molecule and the membrane is modeled to grow discontinuously for simplicity. When the number is increased, the searching activity mentioned in the following is carried out $T$ times. When the search at $2n_{M0}$ is finished, the cell is divided into two daughter cells.

At the initial stage with $n_{M0}$ membrane molecules, the initial ligase L tries up to $T$ times search for S1 and S2 to meet the conditions (i) and (ii). The candidate oligo-nucleotides are randomly taken into the cell from the environment which is assumed to be a random pool with oligomers, proto-ATPs and so on. At the same time oligo-peptides are also randomly taken into according to the relation (1). If the search fails, the simulation go to the next stage by increasing $n_M$, namely expanding the cell. If the search succeeds, the ligase L and the substrates S1 and S2 form a complex. The complex keeps the binding between L and S1-S2 until it meets with the helicase H. If H is present in the cell, H immediately separates L and S1-S2. H is at once formed, with the probability $p$, if both D1 and D2 are present in the cell. After the separation the initial ligase L regains the ligation activity and repeats the search. At the same time the combined S1-S2 20-mer becomes a new ligase and tries up to $T-T_L$ times search for the substrates to meet the conditions (i) and (ii). Here $T_L$ is the number of the trial for L to find S1-S2 substrate. Within the allowed trials all the ligases can make their descendants and the descendants act as ligases. The number of allowed trials becomes smaller for descendants as mentioned above. In the presence of helicases a ligase is doubled at each ligation reaction.

When the allowed trials are finished, the number of membrane molecules increases by $\Delta n_M$. Then new oligomers are taken into the cell from the environment and new searches up to $T$ times begin. These procedures continue until the cell division.

If more than two ligases exist in a cell, a ligase can be a substrate for the other ligases. Thus 10-mer and 20-mer can be combined into 30-mer. Moreover 20-mer and 20-mer can be combined into 40-mer. For such ligations the condition (ii) is modified as follows, while (i) is unchanged. (ii)': the candidate 6 continued bases to meet ligase's 6 end-bases can be chosen from 1-st to 16-th bases from the end of the substrate 20-mer. The relation (1) is also modified to account 30-mers and 40-mers. Such a scheme to develop longer ligases has been also discussed in the literature (Cheng and Unrau, 2010).

## 3. Simulation

We have started the simulation with a cell which contains a 20-mer ligase and taken the data at $n_M = 2n_{M0}$. The data in Fig. 3 shows that typically the number of ligases in a cell increases from 1 to 70 before the cell division and 88% of samples generate 30-mers. The average number of 20-mer ligases in a cell is 48. The average number of 30-mers in a cell is 3.3.

Since our condition (ii) doesn't require complete complementation, our product of the replication is incomplete. The data in Fig. 4 shows the incompleteness of the replication. The complementation is defined by the rate, (the number of the Watson-Crick pairs except the functional part) / (the number of the bases of the ligase except the functional part), where the latter number is 16. The number of the Watson-Crick pairs is measured between the original ligase and the $2n$-th generation ligases or between the ideal 2-nd generation ligase and the ($2n+1$)-th generation ligases. Here $n = 1,2,3,\cdots$. The ($n+1$)-th generation ligases are produced by the $n$-th generation ligases. The ideal 2-nd generation ligase is defined to be completely complementary, except the functional part, to the original 1-st generation ligase. Although the replication is incomplete, the diversity in the 20-mers arises.

In Fig. 5 and Fig. 6 the time interval $T$ of the membrane expansion is changed. As naturally expected, the larger $T$ leads to the more ligations. However, we cannot estimate the value of $T$ which is relevant to experiments.

In Fig. 3, Fig. 5 and Fig. 6 we have confirmed the production of 30-mers. They are longer than the original ligase so that we have obtained more complex polymers in an easy way. This gain of the complexity is essential for the evolution.

The 30-mers also act as ligases. Then we have performed the simulation starting with a cell which contains one 30-mer ligase. Such a simulation corresponds to the cell activity of the daughter cells. The daughter cells also contain 20-mers succeeded from their mother cells. The 20-mer becomes the substrate for the 30-mer ligase. The 30-mer is partitioned into two parts. The longer part is from 5'-end to 20-th base and the shorter part is from 21-st base to 3'-end. The functional bases, G-A-A-G, are located from 19-th to 22-nd. The longer part is assumed to bind with any 20-mers which have G-A bases at 5'-end. The shorter part binds either 10-mer or 20-mer according to the same condition as mentioned above. In Fig. 7, Fig. 8 and Fig. 9 we have confirmed that the 30-mers act as ligases and that longer 40-mers are generated. The latter result show the easy way to gain complexity. The descendants of the daughter cells will gain more complexity easily. The increase of the complexity is necessary for the evolution.

The above simulations have been carried out with $p = 1$ where $p$ is the probability for the helicase formation in the presence of helicase domains D1 and D2. As naturally expected, the larger $p$ leads to the more ligations. However, we cannot estimate the value of $p$ which is relevant to experiments. In Fig. 10 we show the results for smaller $p$.

## 4. Conclusion

We have proposed a mechanism of the replication of proto-RNAs in oligomer world. The replication is sustained by the ligase-helicase cycle. The ligase can be implemented at the level of oligomers, while a polymerase which assembles RNAs from monomers is inevitably more complex polymer even in vitro. The helicase can be also implemented at the level of oligomers, while the present-day helicase is a more complex molecular motor.

The template of the replication is the ligase and the product of the ligation is not completely complementary to the template. As the result the products acquire diversity. At the same time the products can become longer than the template. As the result they acquire complexity. Both diversity and complexity are necessary for evolution.

Several processes needed to sustain our system have not been implemented for simplicity. Such processes are the motion of the helicase, the circulation of proto-ATPs and the reaction to produce membrane molecules. The implementation of such processes can be included in our forthcoming models.

Several parameters in our simulation lack quantitative grounds compared with experiments. To establish the connection with experiments is a future task.

**Figures and Captions**

**Fig. 3.** Histograms for 10,000 samples where every sample cell contains the common original ligase, whose base sequence is CACACGUUGAAGGUUGGCCG, but the oligomers in the environment are generated randomly. A sample starts with $n_{M0} = 2{,}500$ and the membrane expands by $\Delta n_M = 50$. The oligo-peptide ratio at the expansion is chosen as $R = 0.3$. The interval time of the expansion is set as $T = 500$. The helicase-domain ratio is chosen as $D = 0.02$ and the probability of the formation of a helicase as $p = 1$. (Left) Every sample is classified by the number of ligases contained in the cell at $n_M = 2n_{M0}$. (Right) Every sample is classified by the number of 30-mers contained in the cell at $n_M = 2n_{M0}$.

**Fig. 4.** Histograms for 70 ligases in a cell at $n_M = 2n_{M0}$ obtained by the same simulation as Fig. 3. The ligases are classified by the complementation.

**Fig. 5.** Histograms for 10,000 samples obtained by almost the same simulation as Fig. 3. Only the interval time of the expansion is different as $T = 250$.

**Fig. 6.** Histograms for 10,000 samples obtained by almost the same simulation as Fig. 3. Only the interval time of the expansion is different as $T = 50$.

**Fig. 7.** Histograms for 10,000 samples obtained by almost the same simulation as Fig. 3. Only the initial condition is different. The initial cell contains one 30-mer ligase and thirty five 20-mer ligases. Every sample is classified by the number of 30-mers or 40-mers contained in the cell at $n_M = 2n_{M0}$.

**Fig. 8.** Histograms for 10,000 samples obtained by almost the same simulation as Fig. 7. Only the interval time of the expansion is different as $T = 250$.

**Fig. 9.** Histograms for 10,000 samples obtained by almost the same simulation as Fig. 7. Only the interval time of the expansion is different as $T = 50$.

**Fig. 10.** Histograms for 10,000 samples obtained by almost the same simulation as Fig. 7. Only the probability of the formation of a helicase is different as $p = 0.2$ (Left) and $p = 0.004$ (Right).

**Fig. 3** (Left)

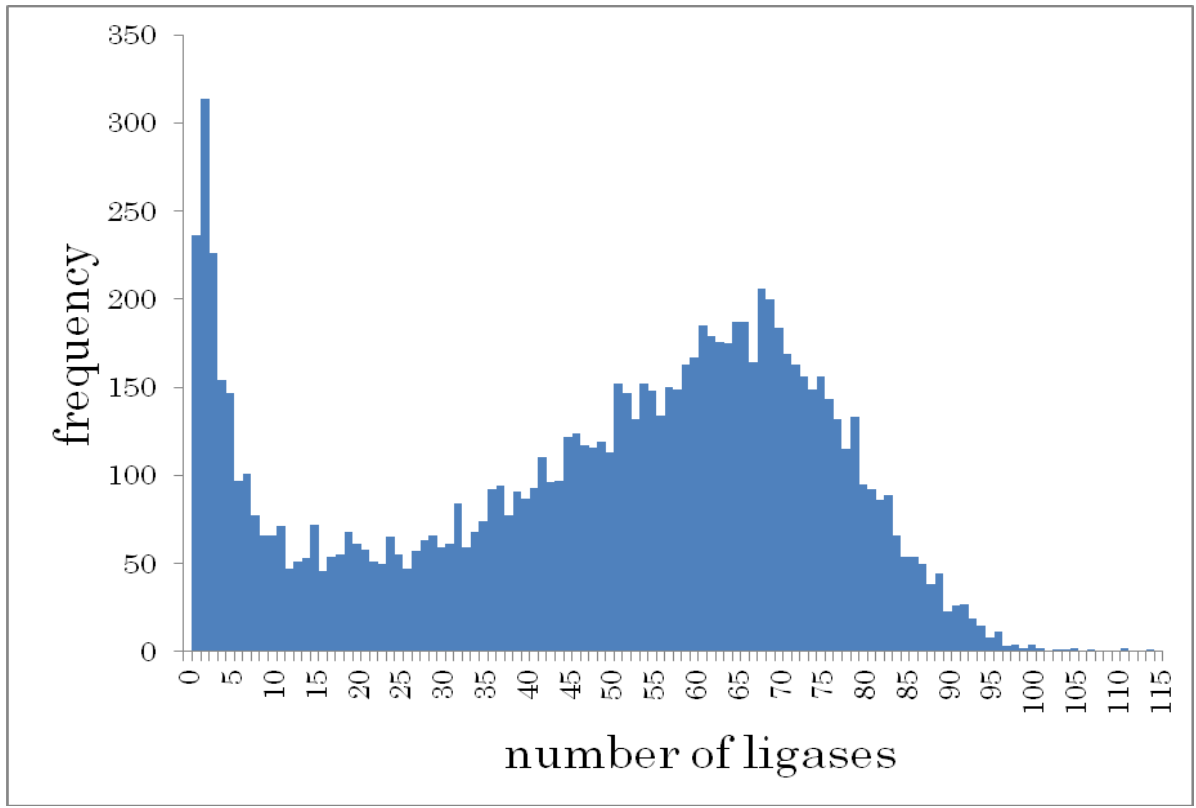

**Fig. 3** (Right)

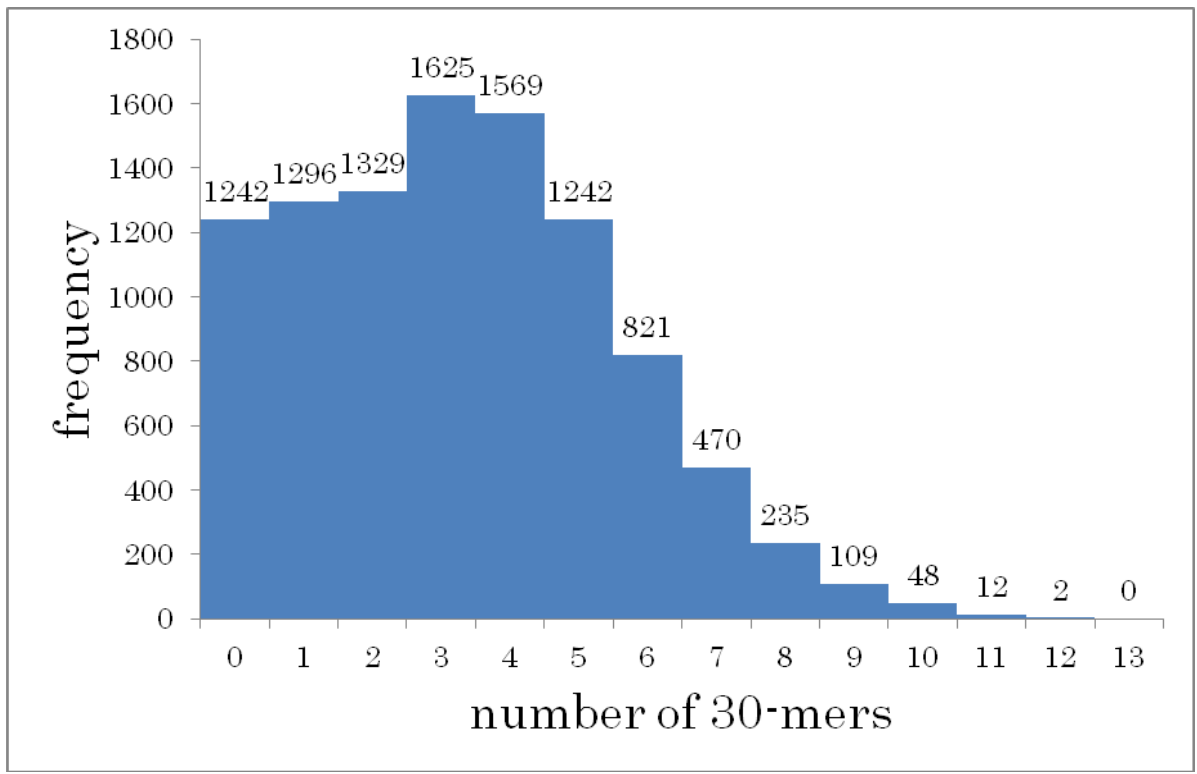

**Fig. 4**

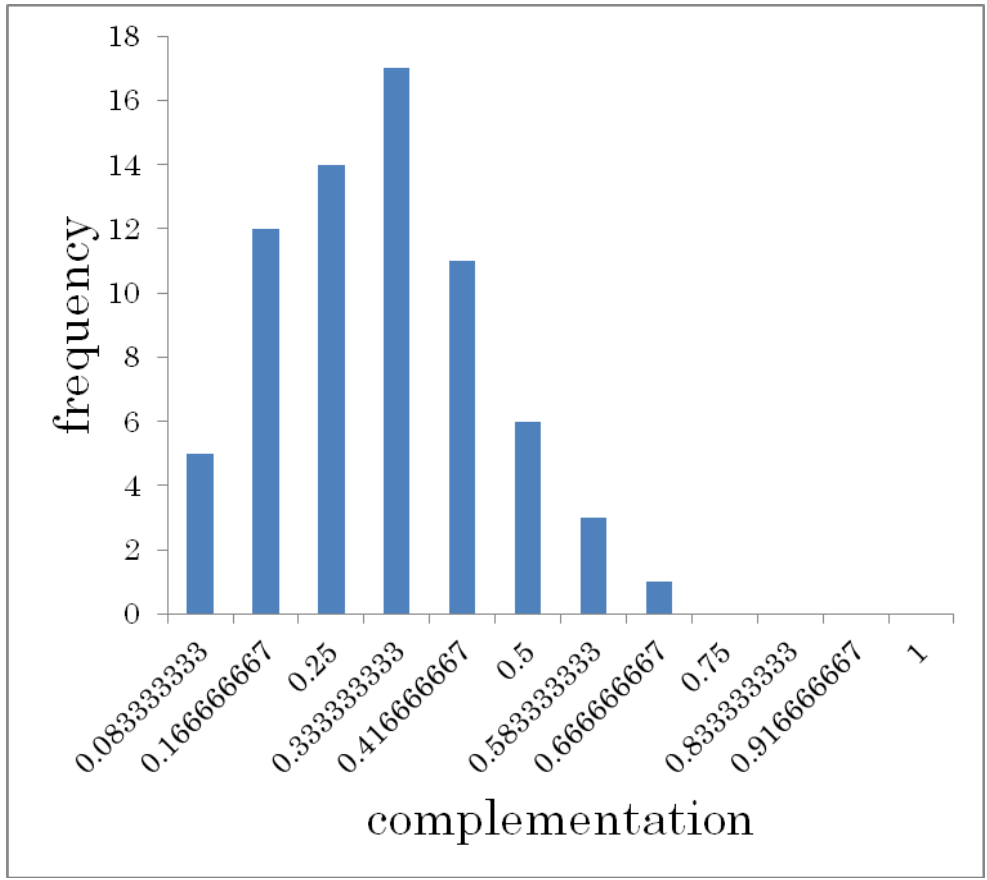

**Fig. 5** (Left)

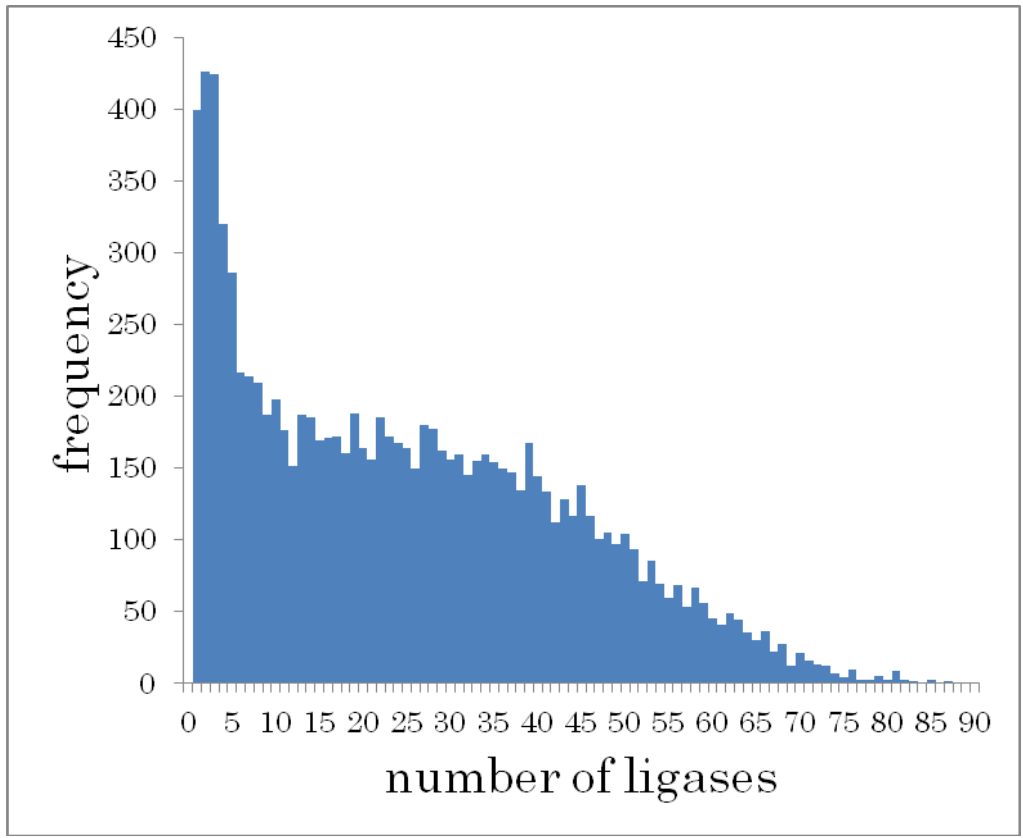

**Fig. 5** (Right)

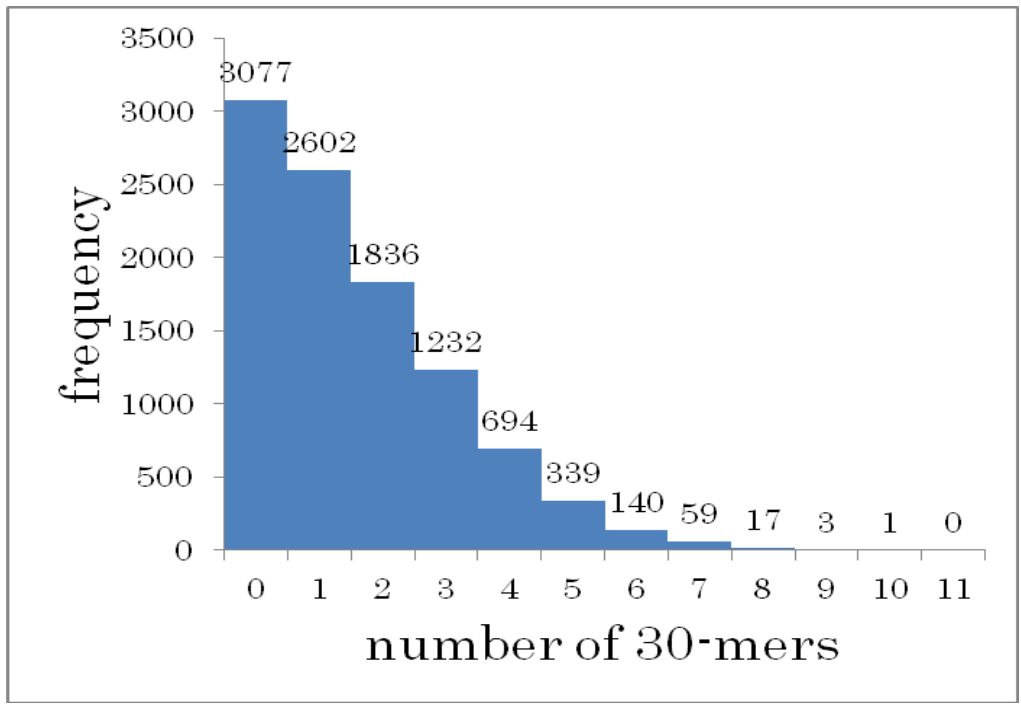

**Fig. 6** (Left)

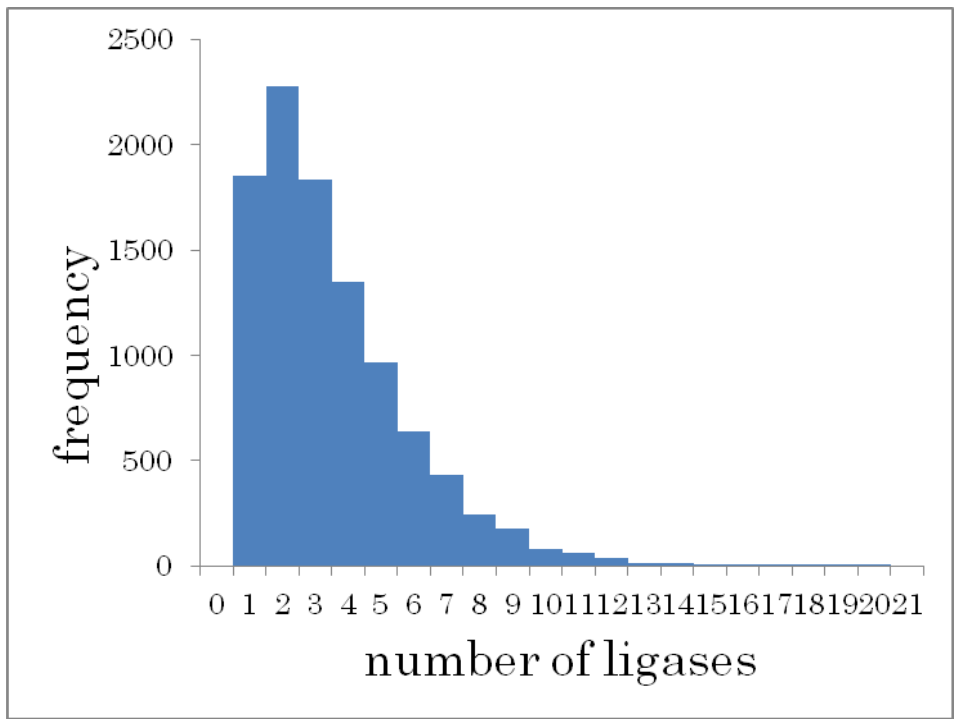

**Fig. 6** (Right)

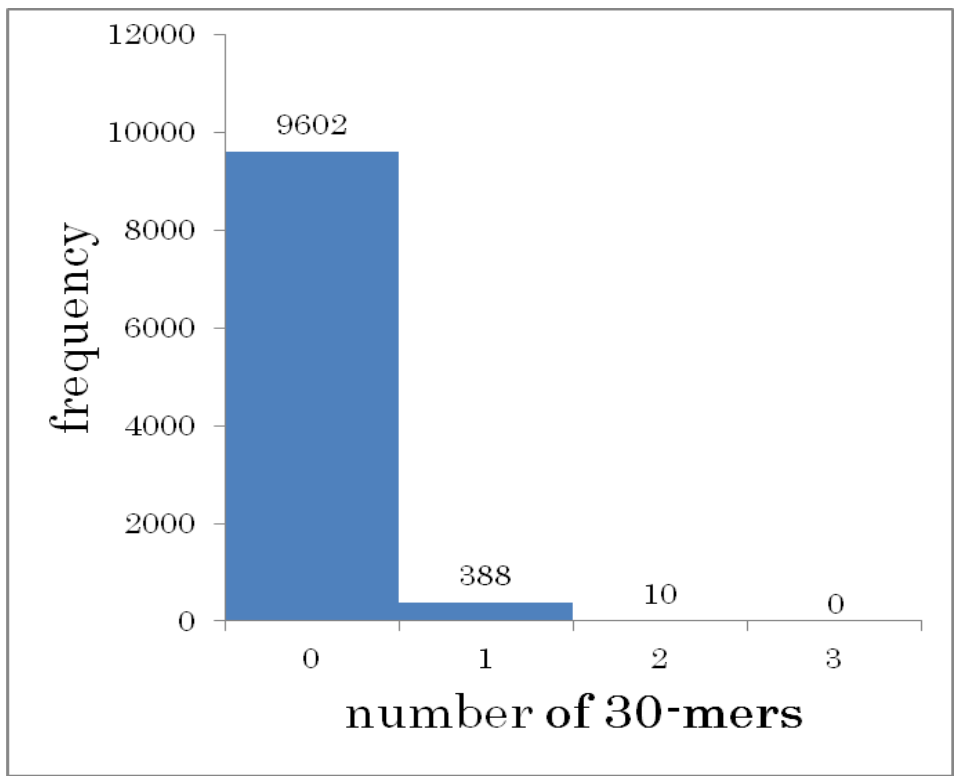

**Fig. 7** (Left)

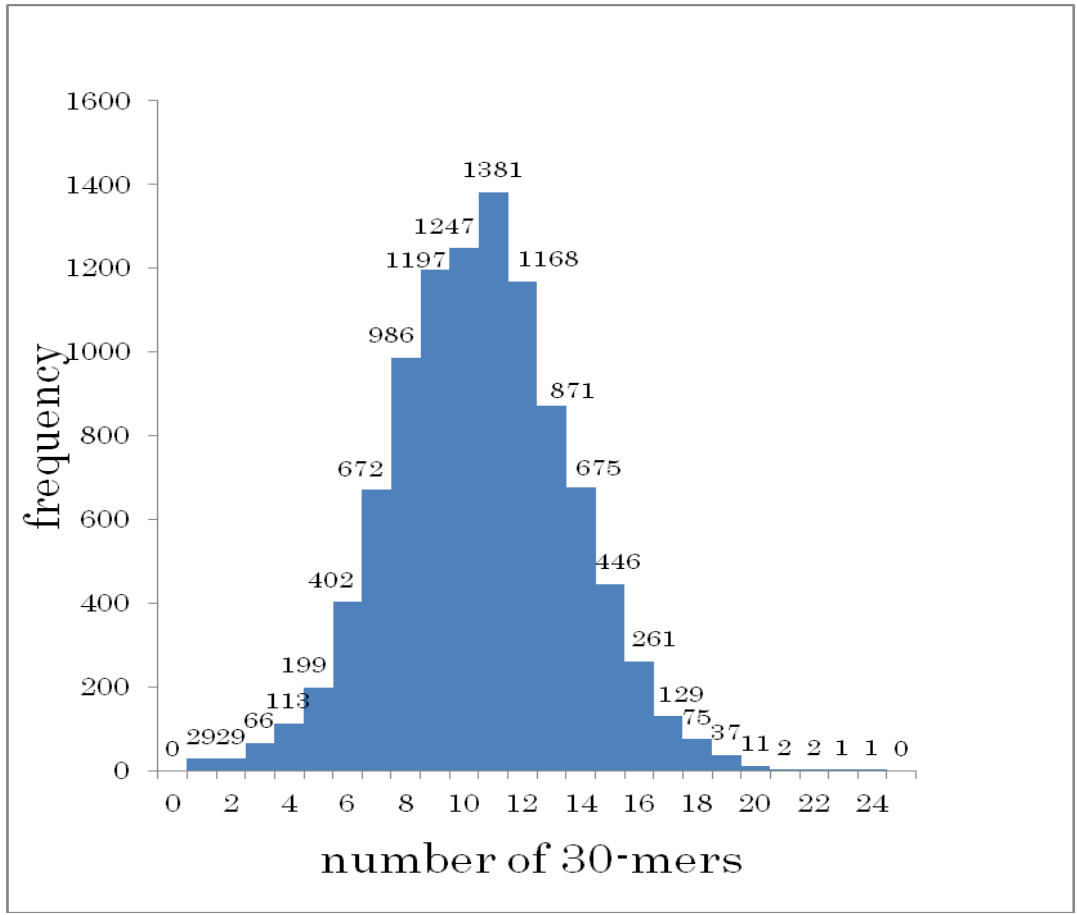

**Fig. 7** (Right)

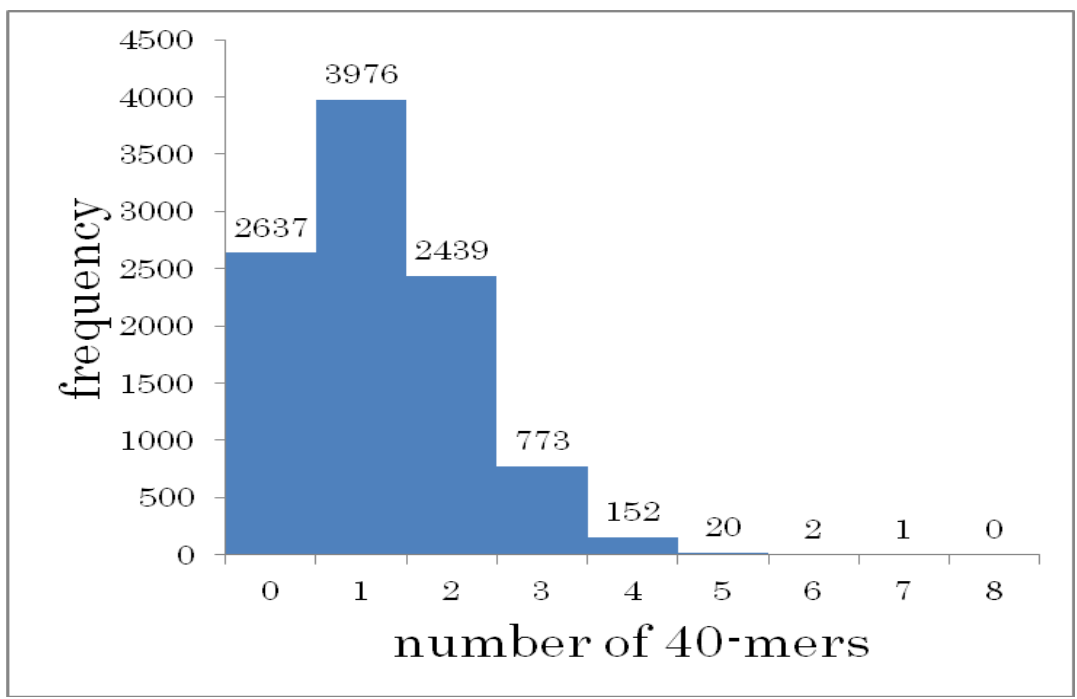

**Fig. 8** (Left)

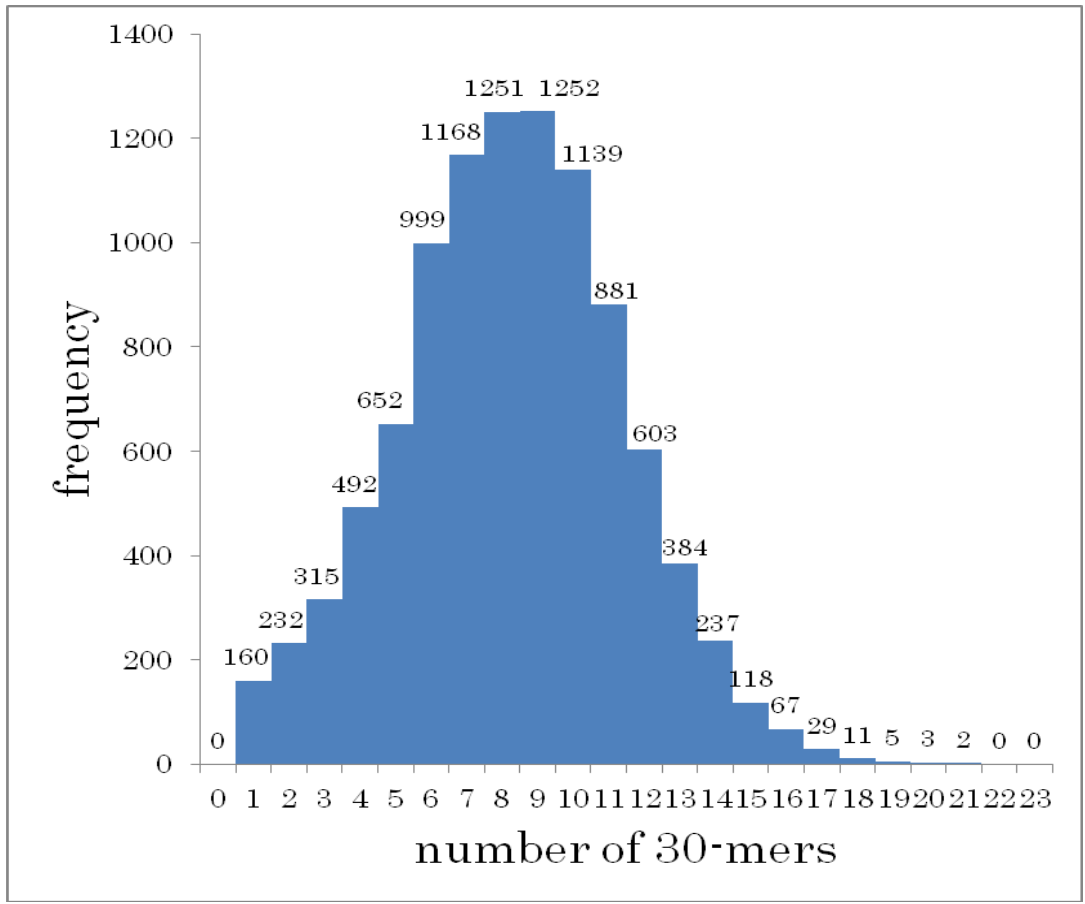

**Fig. 8** (Right)

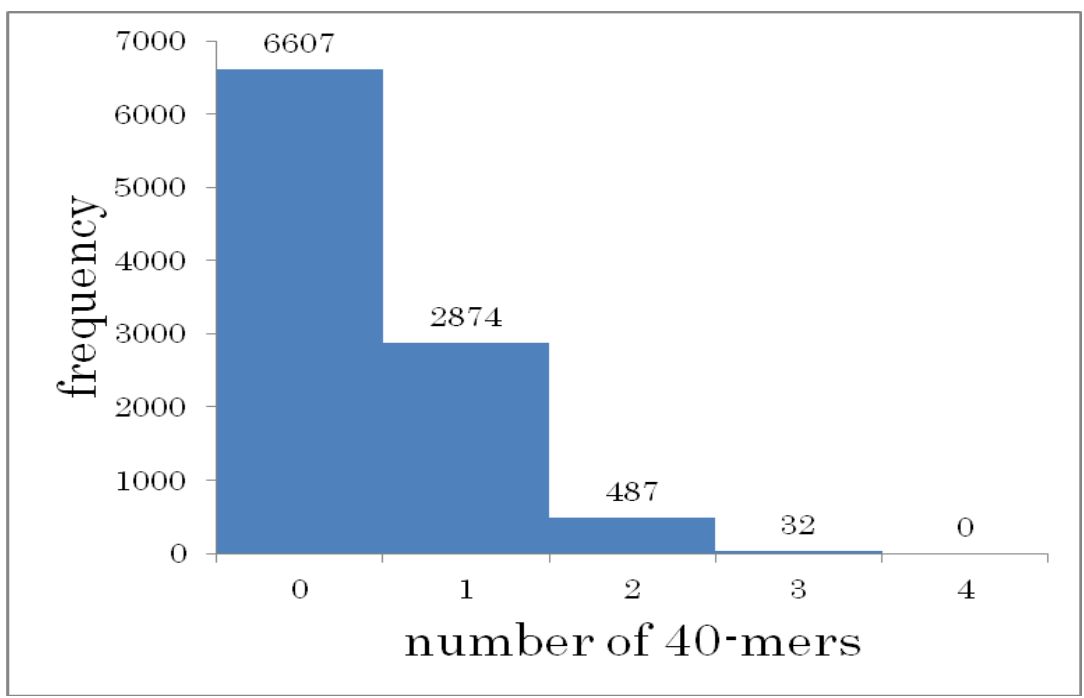

**Fig. 9** (Left)

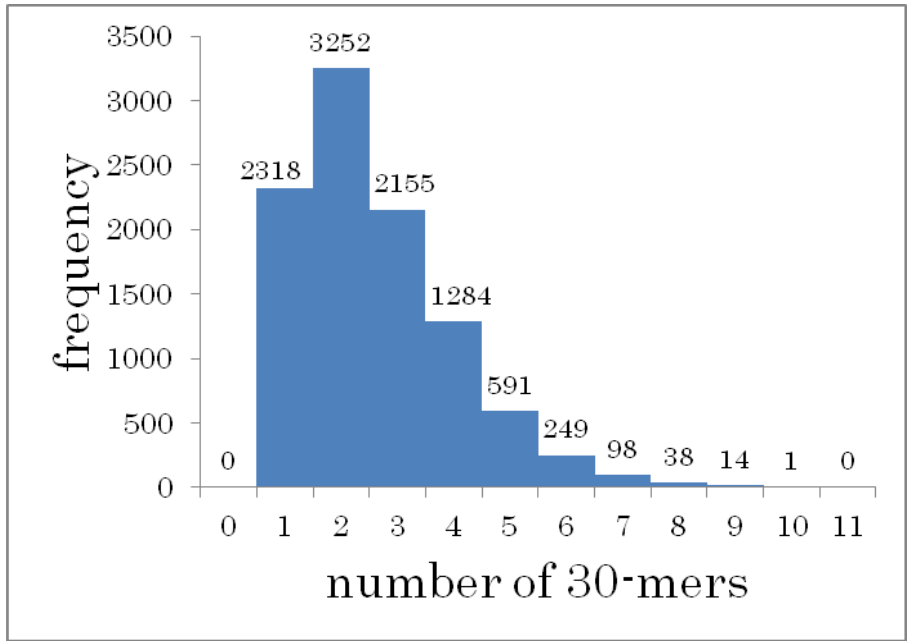

**Fig. 9** (Right)

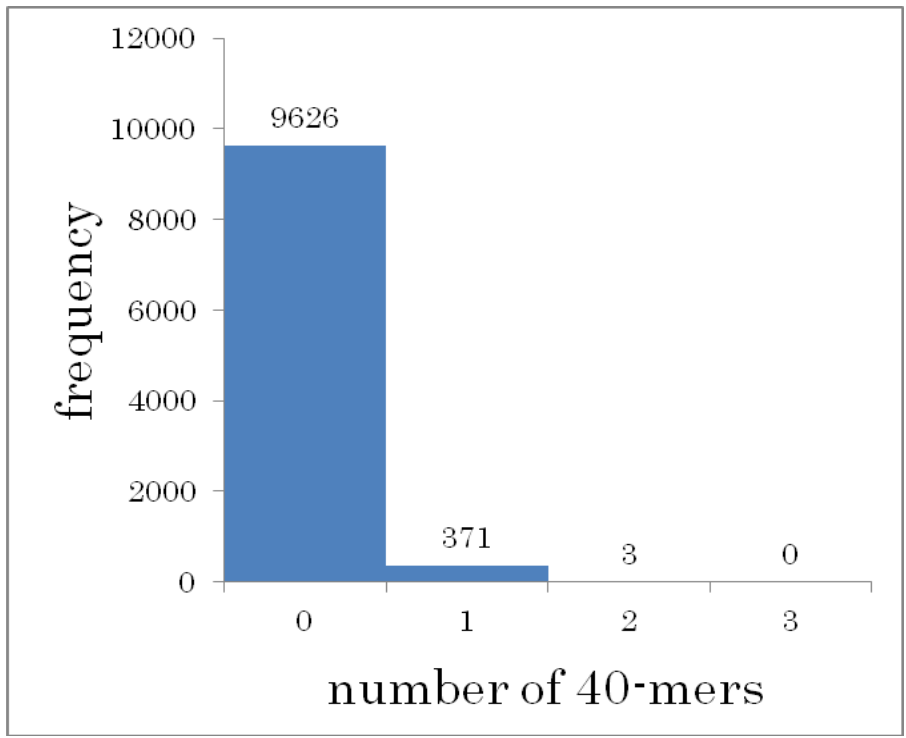

**Fig. 10** (Left)

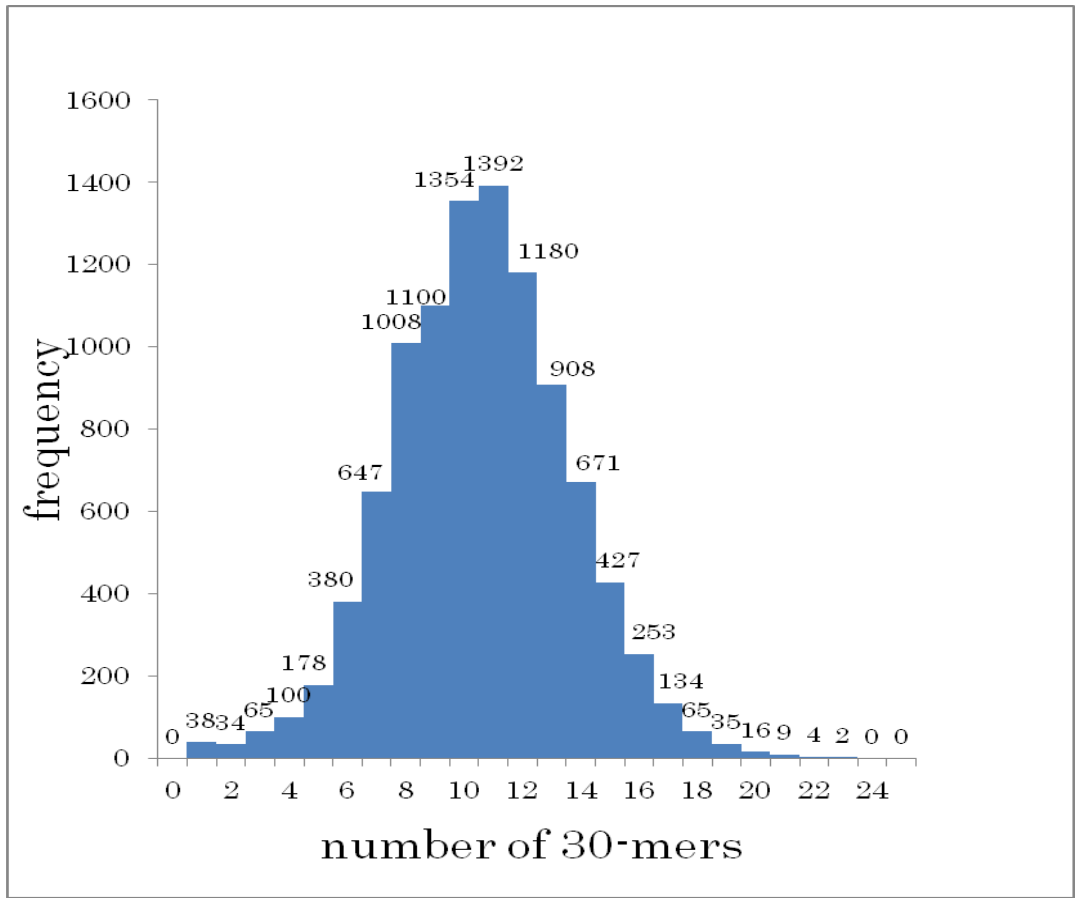

**Fig. 10** (Right)

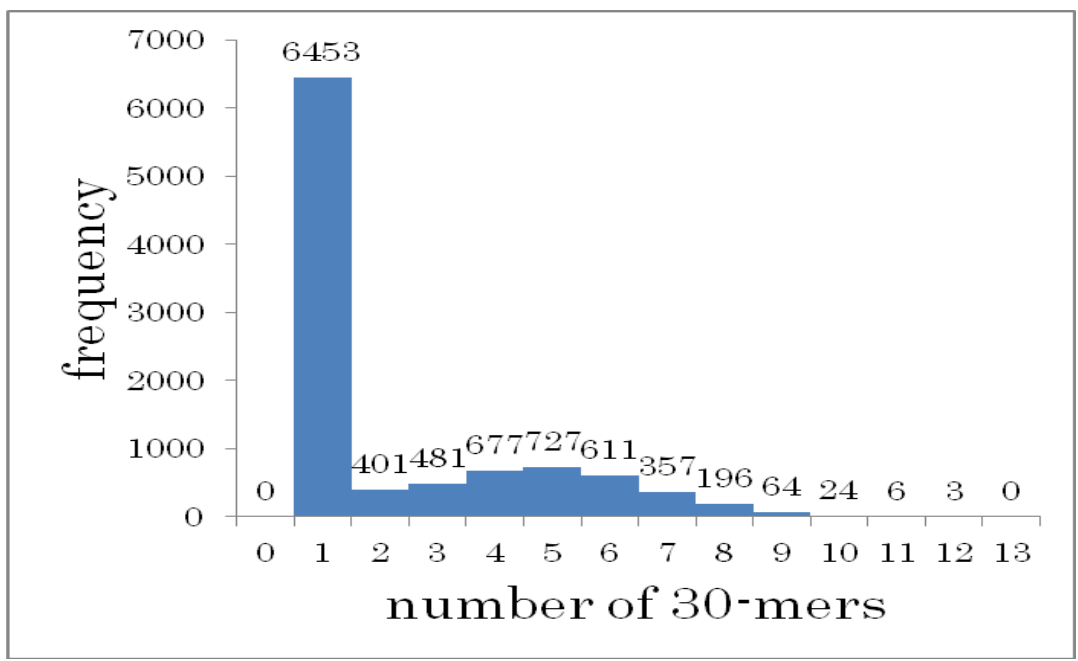